\documentclass{article}

\usepackage{arxiv}

\usepackage[utf8]{inputenc} 
\usepackage[T1]{fontenc}    
\usepackage{booktabs}       
\usepackage{amsfonts}       
\usepackage{nicefrac}       
\usepackage{microtype}      
\usepackage{lipsum}
\PassOptionsToPackage{hyphens}{url}\usepackage{hyperref}

\usepackage{cite}
\usepackage{amsmath,amssymb,amsfonts}
\usepackage{algorithmic}
\usepackage{graphicx}
\usepackage{textcomp}
\usepackage{xcolor}
\usepackage{multirow}
\usepackage[caption=false,font=footnotesize]{subfig}

\title{GIDS: GAN based Intrusion Detection System for In-Vehicle Network}

\author{
  Eunbi Seo \\
  Graduate School of Information Security\\
  Korea University\\
  Seoul, Republic of Korea \\
  \texttt{eunbi@korea.ac.kr} \\
  \And
  Hyun Min Song \\
  Graduate School of Information Security\\
  Korea University\\
  Seoul, Republic of Korea \\
  \texttt{signos@korea.ac.kr} \\
  \And
  Huy Kang Kim \\
  Graduate School of Information Security\\
  Korea University\\
  Seoul, Republic of Korea \\
  \texttt{cenda@korea.ac.kr} \\  
}

\begin{document}
\maketitle

\begin{abstract}
A Controller Area Network (CAN) bus in the vehicles is an efficient standard bus enabling communication between all Electronic Control Units (ECU). However, CAN bus is not enough to protect itself because of lack of security features. To detect suspicious network connections effectively, the intrusion detection system (IDS) is strongly required. Unlike the traditional IDS for Internet, there are small number of known attack signatures for vehicle networks. Also, IDS for vehicle requires high accuracy because any false-positive error can seriously affect the safety of the driver. To solve this problem, we propose a novel IDS model for in-vehicle networks, \texttt{GIDS} (GAN based Intrusion Detection System) using deep-learning model, Generative Adversarial Nets. \texttt{GIDS} can learn to detect unknown attacks using only normal data. As experiment result, \texttt{GIDS} shows high detection accuracy for four unknown attacks.
\end{abstract}


\section{Introduction}
\label{intro}
The advances in the automotive technology have brought great convenience to driver's life. However, as V2X technology enables interactions with vehicles and everything from outside (e.g., vehicles, infrastructure), security threats on ECU of vehicles become higher. Therefore, we need to develop a security system to mitigate the various risks of the vehicle. In particular, intrusion detection system (IDS) for in-vehicle network is required to protect all of the ECUs and related equipment in the vehicle from emerging threats. 

Controller Area Network (CAN) is a standard of the bus system for in-vehicle network and provides efficient communication between ECUs. CAN bus is a reliable and economical serial bus for the in-vehicle network. However, because it uses a broadcast communication without authentication, attackers can access CAN bus easily, and it causes severe risk. For example, an adversary could inject a malicious packet in CAN bus via a vulnerability at one of the numerous external interfaces
Also, many modern cars which have a communication module for infotainment service can be exposed to the attacks via Over-The-Air (OTA) update module. These attacks could result in not only serious malfunctions of the vehicle but also threats to the safety of drivers. 

IDS is the best way to detect and respond known and unknown attacks of today because it can continuously monitor the in-vehicle system and detect suspicious network events generated by ECUs in real time. Recently, there has been some research for IDS to detect attacks targeted on the vehicles. For example, Song \textit{et al.} proposed a detection model based on time interval analysis of CAN data\cite{song2016intrusion}, and Lee \textit{et al.} presented a method to detect intrusion by monitoring the time interval of the request and response of CAN data\cite{leeotids}. 

Although these models are lightweight and efficient, they have some limitations. When in-vehicle environments are changed, it can require a lot of updates. Also, targets to be detected may be limited since specific attacks are reflected when constructing detection system. If IDS be leaked to the attacker, the attacker can manipulate and avoid detection. To solve these problems, we propose \texttt{GIDS} (Generative Adversarial Nets based Intrusion Detection System) which has the following characteristics: expandability, effectiveness, and security. 

\begin{enumerate} 
	\item \textbf{Expandability}: \texttt{GIDS} maintains consistent detection methodology even if in-vehicle environments are changed. It requires only one training process.
	\item \textbf{Effectiveness}: Because \texttt{GIDS} can be trained using only normal data, it \texttt{GIDS} can detect intrusions without being limited to specific types of attacks. Thus, \texttt{GIDS} is likely to detect unknown attacks not used in the implementation process of the IDS.
	\item \textbf{Security}: \texttt{GIDS} is one of the deep-learning model which has the characteristic of black-box. Thus,it is difficult for an attacker to manipulate internal structure of detection system. 
\end{enumerate} 
\subsection{Organization of This Paper}

We introduced in-vehicle networks and IDS for in-vehicle network in \textsection\ref{intro}. The rest of the paper is organized as follows. \textsection\ref{section2} presents the recent researches. We introduce our IDS, \texttt{GIDS} in \textsection\ref{section3}. In \textsection\ref{section4}, we describe the result of the experiment and discuss the experiment result. Finally we conclude the paper in \textsection\ref{conclusion}. 

\section{Related works}
\label{section2}
The early research for anomaly detection of the in-vehicle system was introduced by Hoppe \textit{et al.}\cite{hoppe2008security}. He presented three selected characteristics as patterns available for anomaly detection that include the recognition of an increased frequency of cyclic CAN messages, the observation of low-level communication characteristics, and the identification of obvious misuse of message IDs. M{\"u}ter \textit{et al.} proposed an anomaly detection based entropy\cite{muter2011entropy}. Marchetti \textit{et al.} analyzed and identified anomalies in the sequence of CAN\cite{marchetti2017anomaly}. The proposed model features low memory and computational footprints. SALMAN \textit{et al.} proposed a software-based light-weight IDS and two anomaly-based algorithms based on message cycle time analysis and plausibility analysis of messages\cite{salmandesign}. It contributed to more advanced research in the field of IDS for in-vehicle networks. 

Many security research in various fields has adopted deep-learning methods for IDS. For example, Zhang \textit{et al.} presented a deep-learning method to detect Web attacks by using the specially designed CNN\cite{zhang2017deep}. The method is based on analyzing the HTTP request packets, to which only some preprocessing is needed whereas the tedious feature extraction is done by the CNN itself. Recently, Generative Adversarial Nets (GAN) was adopted to not only image generation but also other research like anomaly detection. Schlegl \textit{et al.} proposed AnoGAN, a deep convolutional generative adversarial network to learn a manifold of normal anatomical variability. The model demonstrated that the approach correctly identifies anomalous images, such as images containing retinal fluid\cite{schlegl2017unsupervised}.

Although various studies using GAN have been published, most of them are focused only on discrimination of image data. GAN could be useful for security such as IDS. However, few works have explored the use of GAN for security of other fields. We developed a GAN based IDS for in-vehicle security and showed high performance on CAN data that is one of the in-vehicle network datasets. We proved expandability, effectiveness, and security of the proposed model for in-vehicle networks.

\section{\texttt{GIDS}: GAN based IDS}
\label{section3}
\subsection{Converting CAN Data to Image}
\begin{figure}[!h] 
	\centering
	\includegraphics[width=0.5\textwidth]{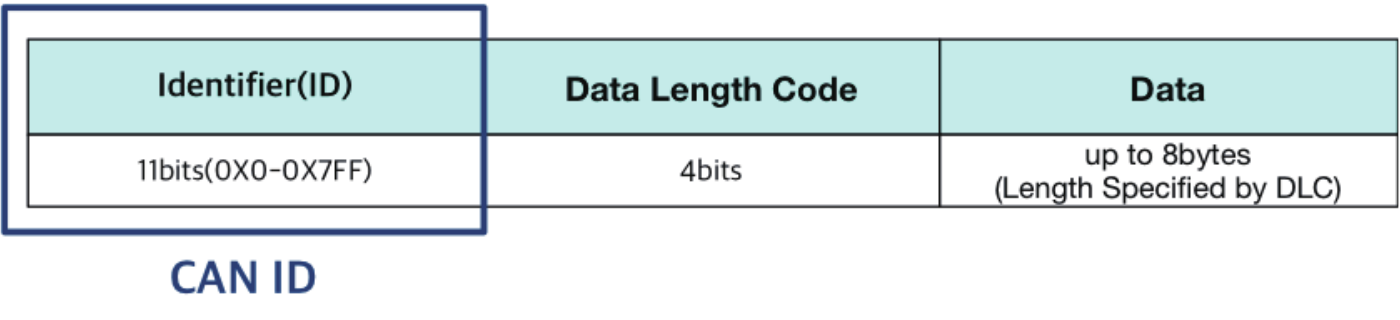}
	\caption{Structure of CAN frame}
	\label{fig:fig1}
\end{figure}

CAN bus supports the ECU to ECU communication. In CAN bus, there are frequent transmissions composed of periodically used CAN messages. ECUs in the vehicle generate about 2,000 CAN data per second to CAN bus. A large amount of real-time CAN data generated by ECUs are must be able to be processed. If all the bits of CAN data are used directly for image conversion, the converted image can be very complex. In the case, \texttt{\texttt{GIDS}} may require a long time not suitable for real-time detection. CAN IDs in CAN data show repetitive patterns and we extracted only patterns of CAN IDs from CAN data for training as in Fig. \ref{fig:fig1}. Also, we converted extracted CAN IDs into a simple image by encoding with one-hot-vector. This method can reduce detection time required for real-time, and improve the performance of IDS.

\begin{figure}[!h] 
	\centering
	\includegraphics[width=0.7\textwidth]{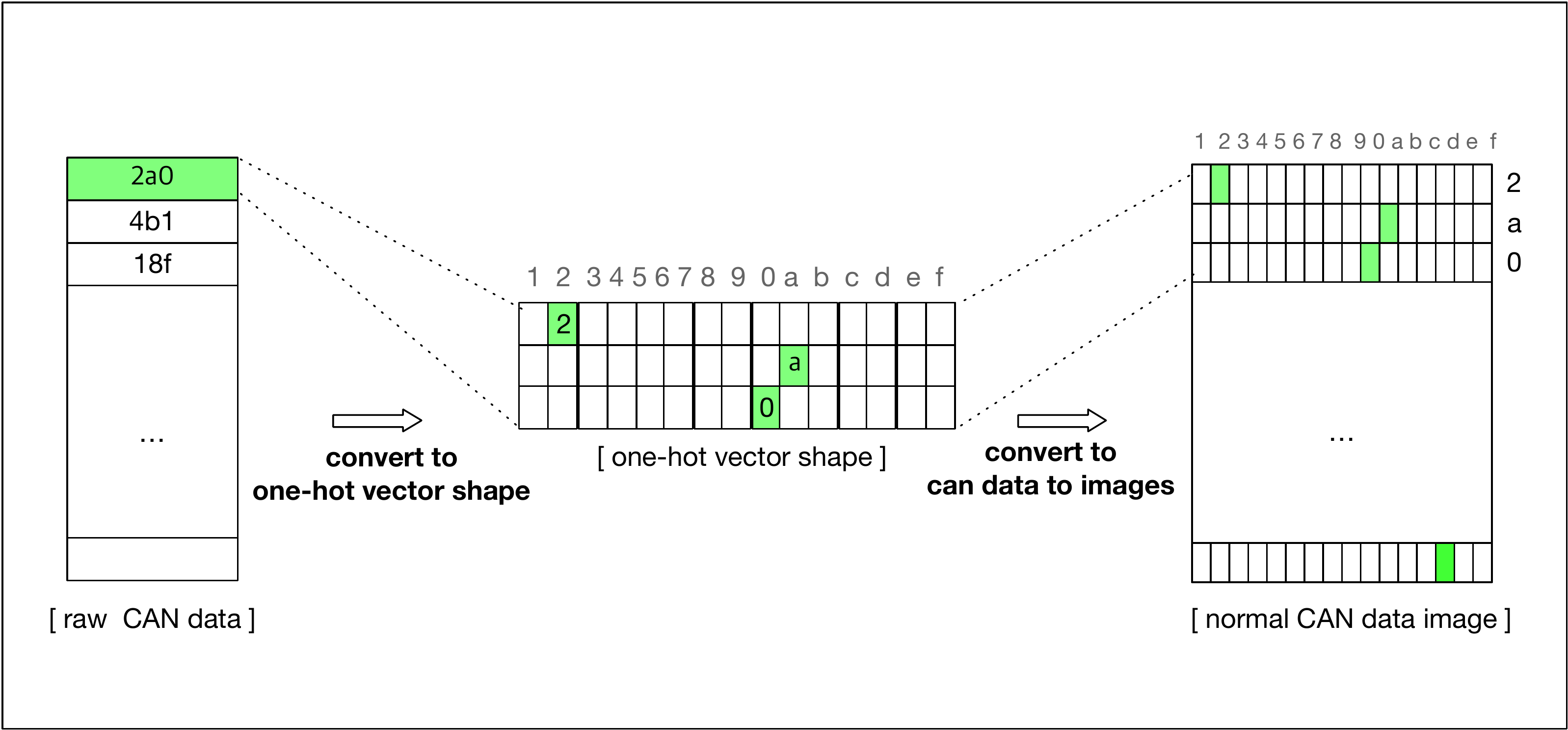}
	\caption{The process of one-hot-vector encoding}
	\label{fig:fig2}
\end{figure}

Fig. \ref{fig:fig2} shows the process of encoding CAN IDs with one-hot-vector. Firstly, because the CAN ID is hexadecimal, each element of the CAN ID such as `2' in `0x2a0' is expressed in a binary form with 16 digits. After that, binary forms of each element of the CAN IDs are encoded to one-hot-vector. Encoding with one-hot-vector makes one of the bits to be 1, and the remaining bits to be all 0. For example, if the element of the CAN ID is `2' in `0x2a0', A one-hot-vector consists of only one bit of the second digit as 1 and the remaining all bits as 0. Finally, a CAN ID of 3-digit such as `0x2a0' is expressed in 16*3 matrix form. For example, if the CAN ID is `0x2a0', it consists of 3 one-hot-vectors such as [0100 ... 000], [0..0100000], and [0..1000000]. We name this matrix as a `CAN image'. 

\begin{figure*}[!h] 
	\centering
	\includegraphics[width=0.9\textwidth]{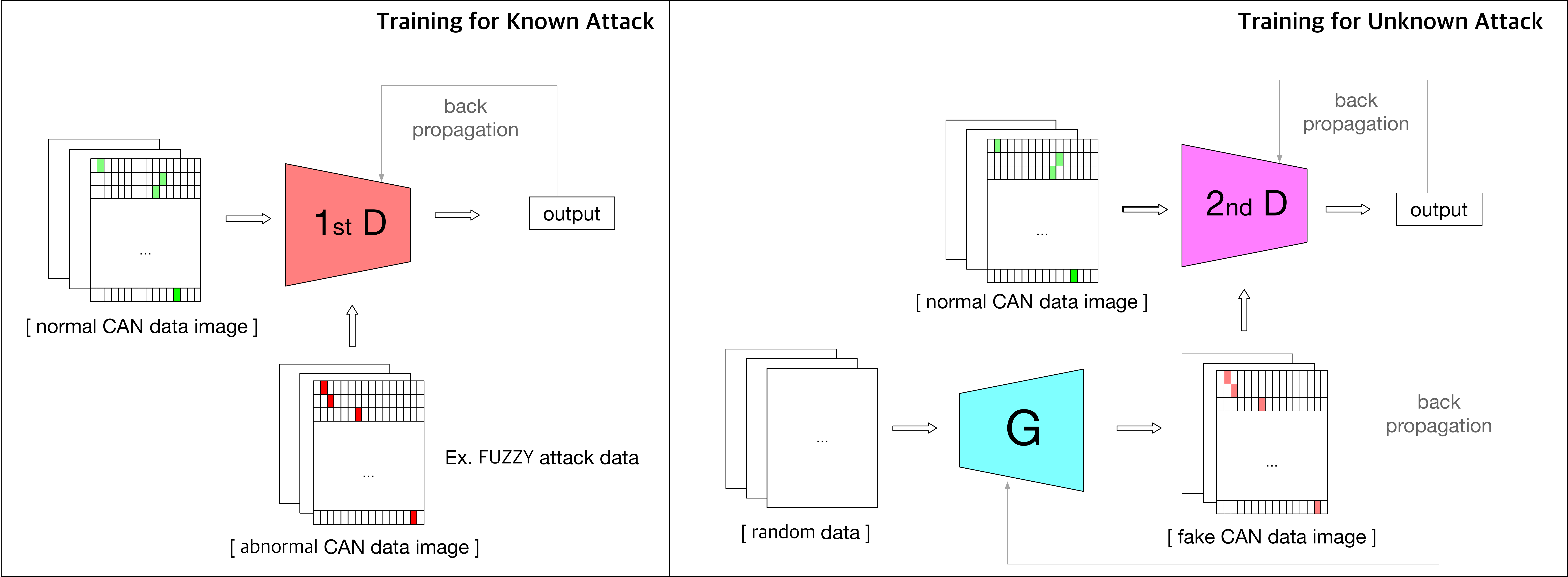}
	\caption{The training process of GIDS}
	\label{fig:fig3}
\end{figure*}

\begin{figure}[h] 
	\centering
	\includegraphics[width=0.7\textwidth]{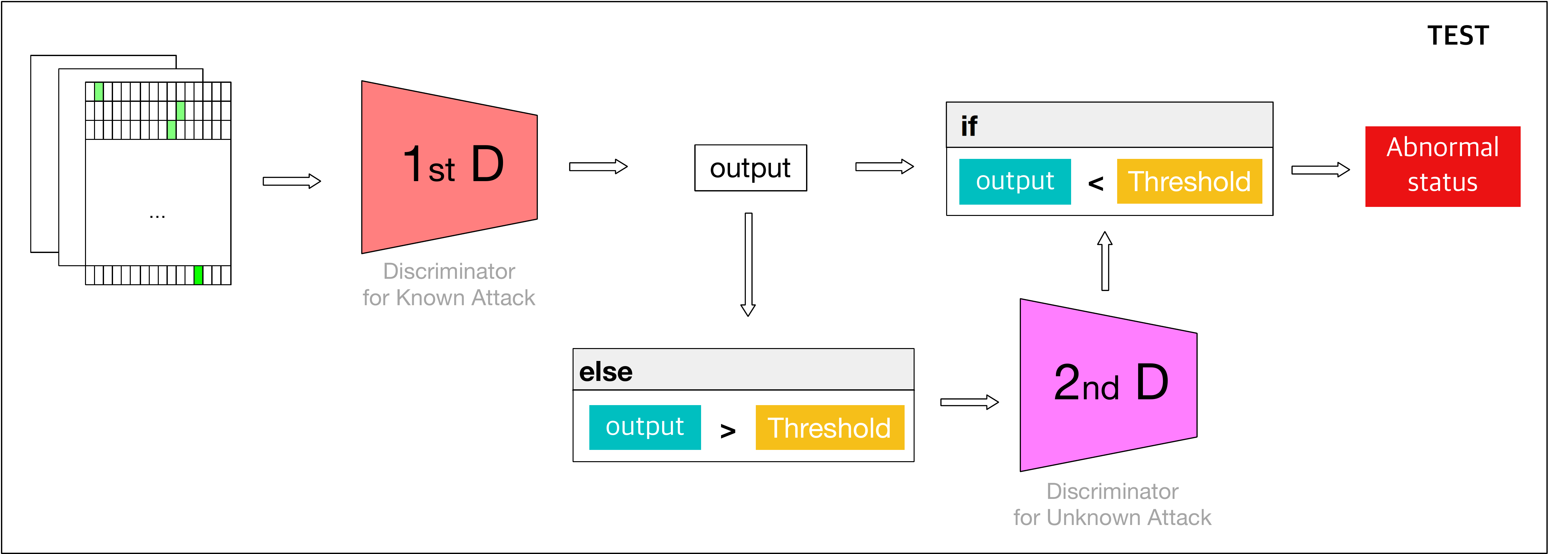}
	\caption{The process of one-hot-vector encoding}
	\label{fig:fig4}
\end{figure}

In this study, we propose GAN based IDS model for the in-vehicle network. We named this model as \texttt{\texttt{GIDS}}. GAN is one of the deep-learning models. GAN is the new framework for estimating generative models via an adversarial process, in which we simultaneously train two models: a generative model G that captures the data distribution, and a discriminative model D that estimates the probability that a sample came from the training data rather than G\cite{goodfellow2014generative}. GAN is often used to generate fake images that are similar to real ones. We focused on the fact and designed our IDS using this fact. \texttt{\texttt{GIDS}} has two discriminative model, the first discriminator and the second discriminator which are trained with the following procedure as shown in Fig. \ref{fig:fig3}.

\begin{enumerate} 
	\item \textbf{Training for known attack}: the First discriminator receives normal CAN images and abnormal CAN images which are extracted from the actual vehicle. Because the first discriminator uses attack data in the training process, the type of attacks that can be detected is likely to be limited to the attacks used for training. 
	\item \textbf{Training for unknown attack}: The generator G and the second discriminator are trained simultaneously by an adversarial process. The generator generates fake images by using random noise. The second discriminator receives normal CAN images and the fake images generated by the generator and estimates the probability that received images are real CAN images. That is, the second discriminator discriminates whether input images are real CAN images or fake images generated by the generator. The generator and the second discriminator compete with each other and increase their performance. In the \texttt{\texttt{GIDS}} model, the second discriminator ultimately win the generator so that the second discriminator can detect even fake images similar to real CAN images.
\end{enumerate} 

\texttt{\texttt{GIDS}} detects attacks of the in-vehicle networks with the following procedure as shown in Fig. \ref{fig:fig4}.

\begin{enumerate} 
	\item \textbf{} The real-time CAN data is encoded with one-hot-vector, and it is converted into CAN images.
	\item \textbf{} The first discriminator receives CAN images and outputs one value which is between 0 and 1.
	\item \textbf{} If output is lower than the threshold, current status is classified as abnormal. 
	(Because the first discriminator is trained for known attacks, unknown attacks are unlikely to be detected in this process.)
	\item \textbf{} If output is higher than the threshold, the corresponding CAN images are received by the second discriminator. As in step 2 and step 3, the second discriminator receives CAN images and outputs one value which is between 0 and 1. 
	\item \textbf{} If output is lower than the threshold, current status is classified as abnormal.
	(Because the second discriminator is trained with only normal data, attack data to be detected are not limited. That is, it may even be possible to detect unknown attacks.)  
\end{enumerate} 

Our goal is to ensure high accuracy for detecting even unpredictable attacks with only normal data. However, if we use only the second discriminator trained with only normal data, the detection accuracy can be lower than when using the first discriminator trained with attack data. Therefore, we combine the first discriminator and the second discriminator, which is able to detect both known attacks and unknown attacks.

\subsection{Design of Neural Networks}
In the chapter, we describe two model structures of the discriminator and the generator in the \texttt{\texttt{GIDS}} model. We measured the detection performance for four combinations of discriminator and generator composed of the convolutional neural network (CNN) and deep neural network (DNN). The neural networks of \texttt{\texttt{GIDS}} was selected as the combination which are shown the best detection performance.

\begin{figure*} [!h] 
	\centering
	\subfloat[Architecture of generator in \texttt{GIDS}]{
		\includegraphics[width=0.5\linewidth]{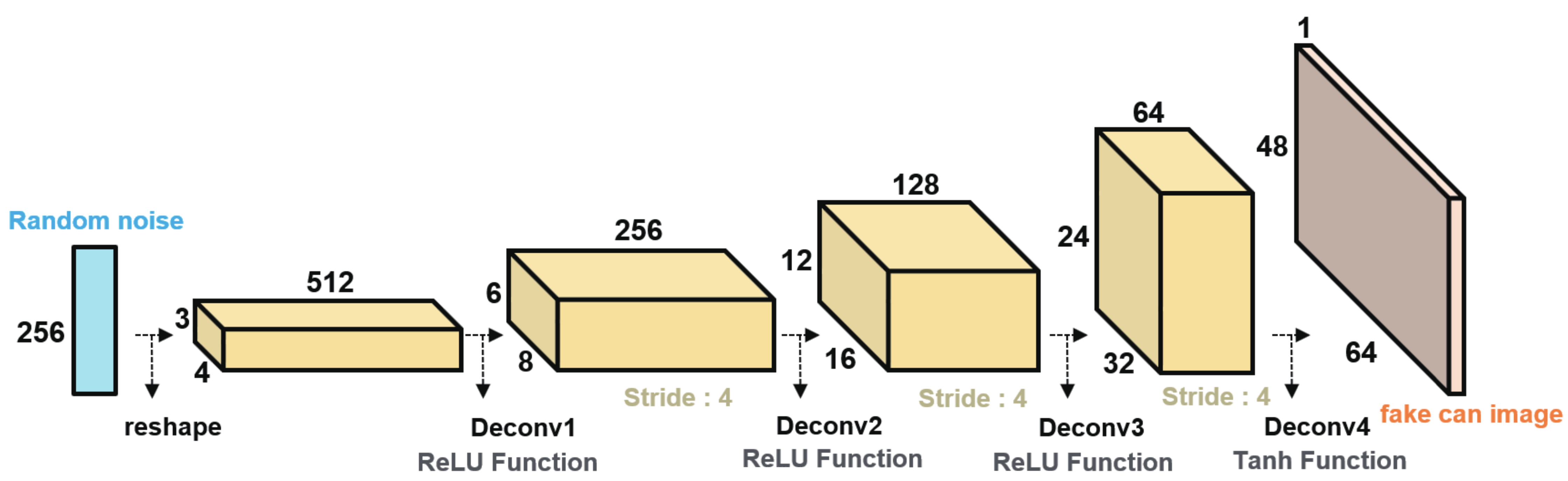}}
	\label{5a}\hfill
	\subfloat[Architecture of discriminator in \texttt{GIDS}]{
		\centering
		\includegraphics[width=0.47\linewidth]{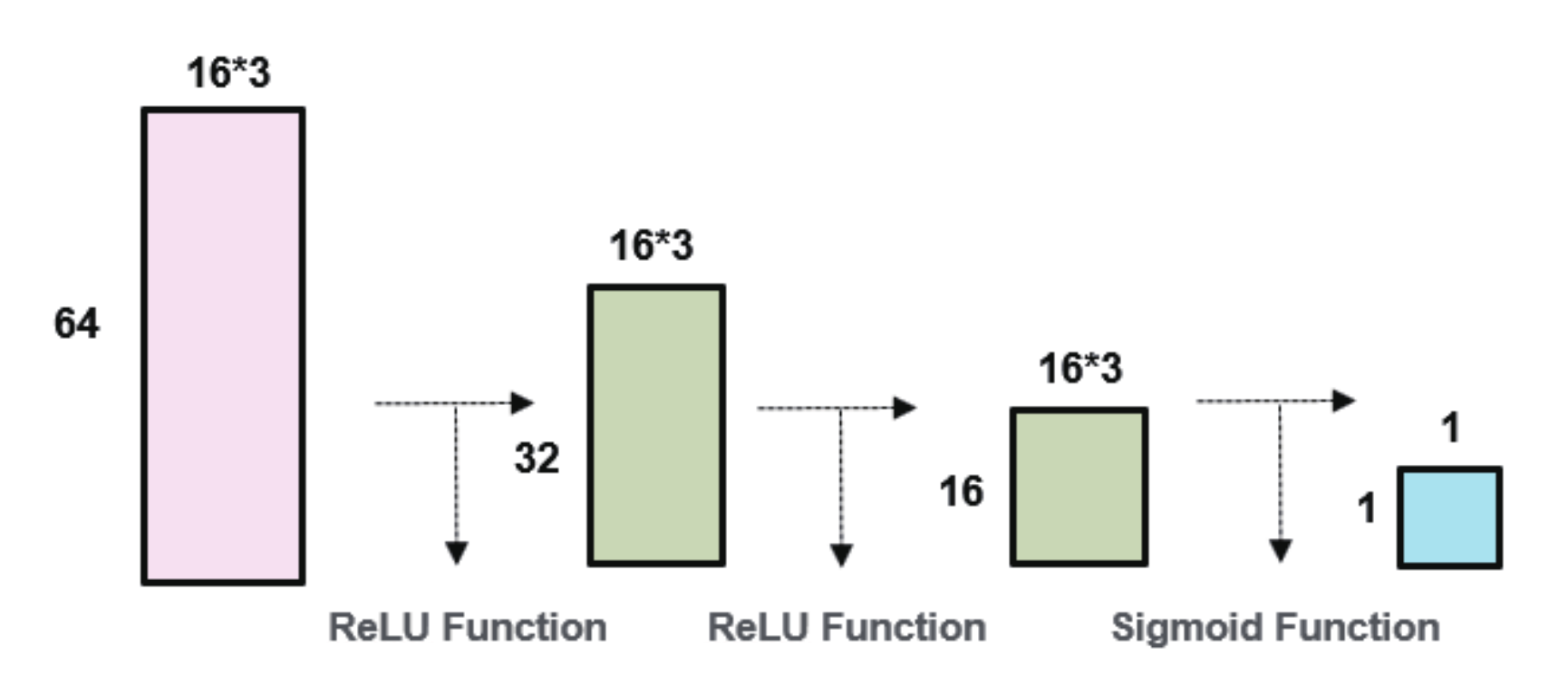}}
	\label{5b}\hfill
	
	\caption{Architecture of \texttt{\texttt{GIDS}}}
	\label{fig:fig5} 
\end{figure*} 

\begin{figure*} [!ht] 
	\centering
	\subfloat[DoS Attack]{
		\includegraphics[width=0.26\linewidth]{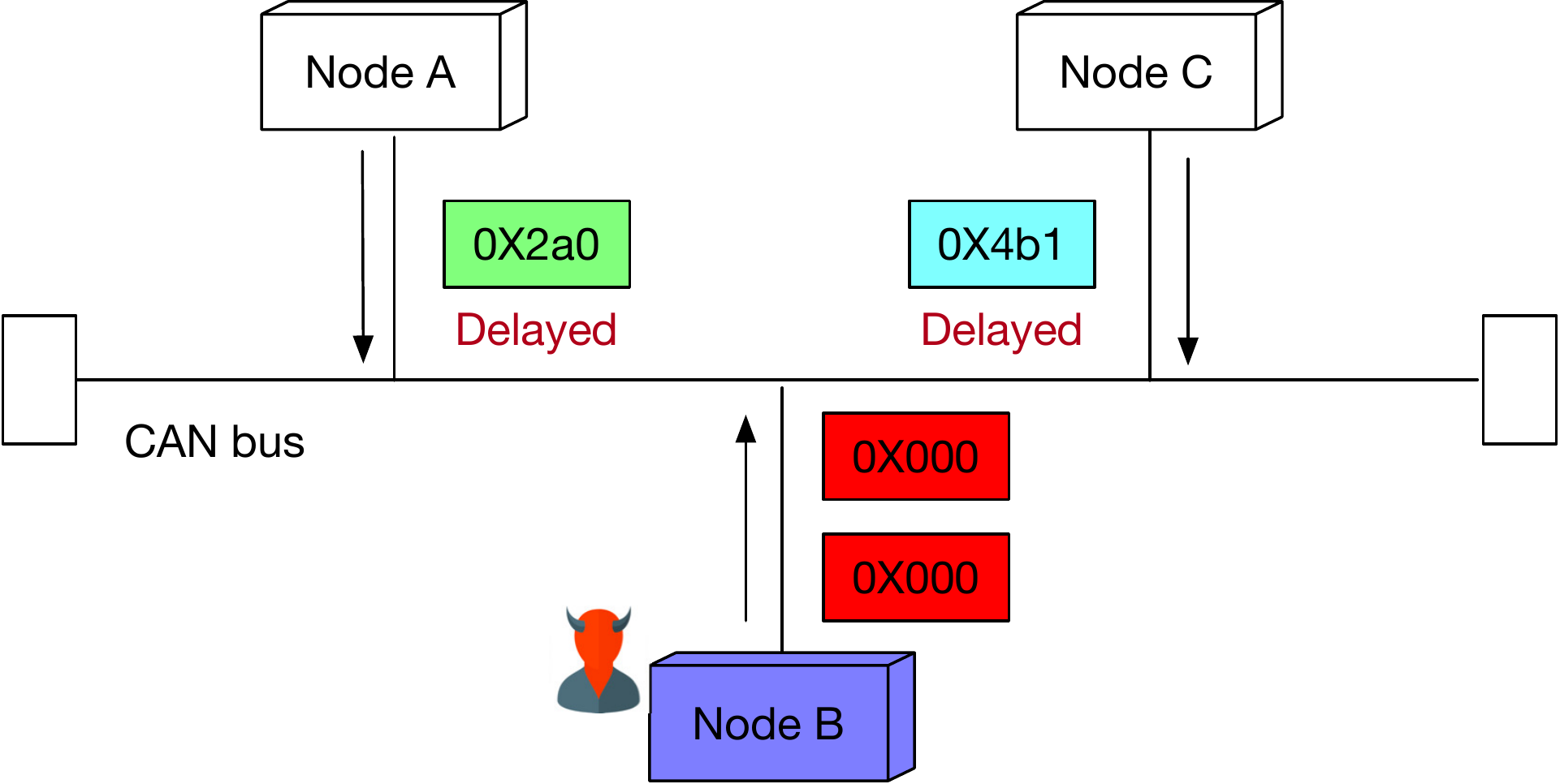}}
	\label{6a}\hfill
	\subfloat[FUZZY Attack]{
		\centering
		\includegraphics[width=0.26\linewidth]{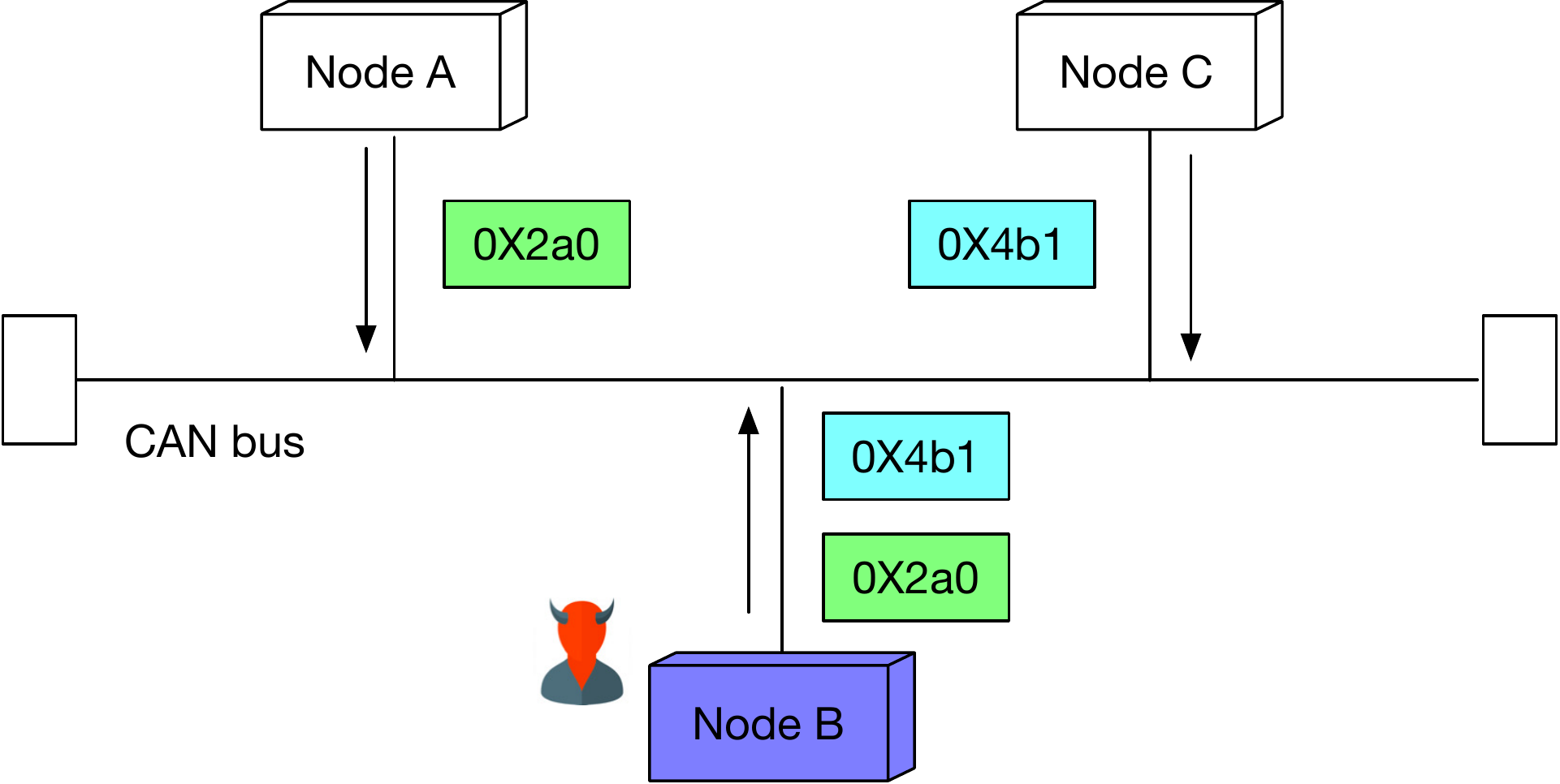}}
	\label{6b}\hfill
	\subfloat[RPM/GEAR Attack]{
		\includegraphics[width=0.26\linewidth]{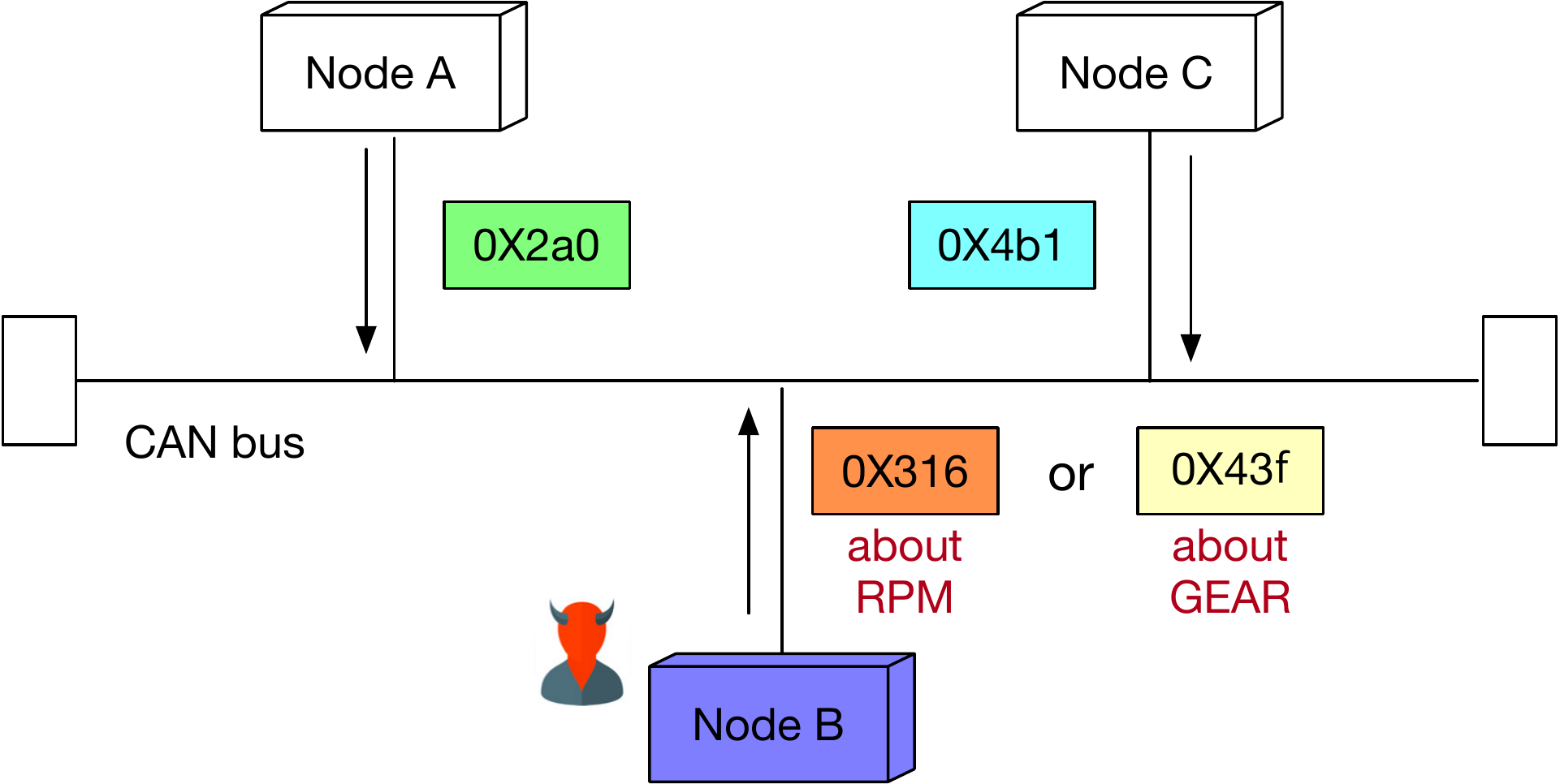}}
	\label{6c}\hfill
	
	\caption{Illustration of DoS, FUZZY and RPM/GEAR attacks}
	\label{fig:fig6} 
\end{figure*}

\begin{enumerate} 
	\item \textbf{Design the discriminator}
	
	The discriminator consists of a deep neural network composed of three layers as shown in Fig. \ref{fig:fig5} (b). The discriminator reduces the dimension of the input data to one output between 0 and 1. Fig. \ref{fig:fig5} (b) shows the process of reduction dimension of the discriminator when the number of CAN IDs is 64. The activation function of each layer is ReLU, and the activation function of the last layer is sigmoid. Finally, the output of the discriminator is used to distinguish between normal status and abnormal status in the in-vehicle network.
	\item \textbf{Design the generator}
	The generator consists of a deconvolutional neural network composed of five layers as shown in \ref{fig:fig5} (a). The generator expands the dimension of random noise data to the one image of the same size as the input data of the discriminator. That is, the generator generates a fake image similar to the real CAN image converted from the CAN IDs. ReLU is used by the activation function of each layer, and Tanh is used as activation function of the last layer. The generator and the discriminator calculate the cost through back-propagation reducing the errors between actual answers and outputs of the model.   
\end{enumerate} 

\section{Experiment and Result}
\label{section4}

\subsection{Experiment Environment}
In the Experiment, we use two criteria: Detection rate and accuracy. The detection rate is defined as the proportion of the detected abnormal data accounting for the total abnormal ones. The accuracy is defined as the proportion of data including normal and abnormal to be correctly classified. We tested the \texttt{GIDS} model in the following experiment environment.

\begin{enumerate}
	\item CPU: Intel(R) Xeon(R) CPU E5-1650 v4 @ 3.60GHz 
	\item RAM: 32.0GB
	\item GPU: NVIDIA GeForce GTX 1080
\end{enumerate}

\begin{figure}[!h] 
	\centering
	\includegraphics[width=0.6\textwidth]{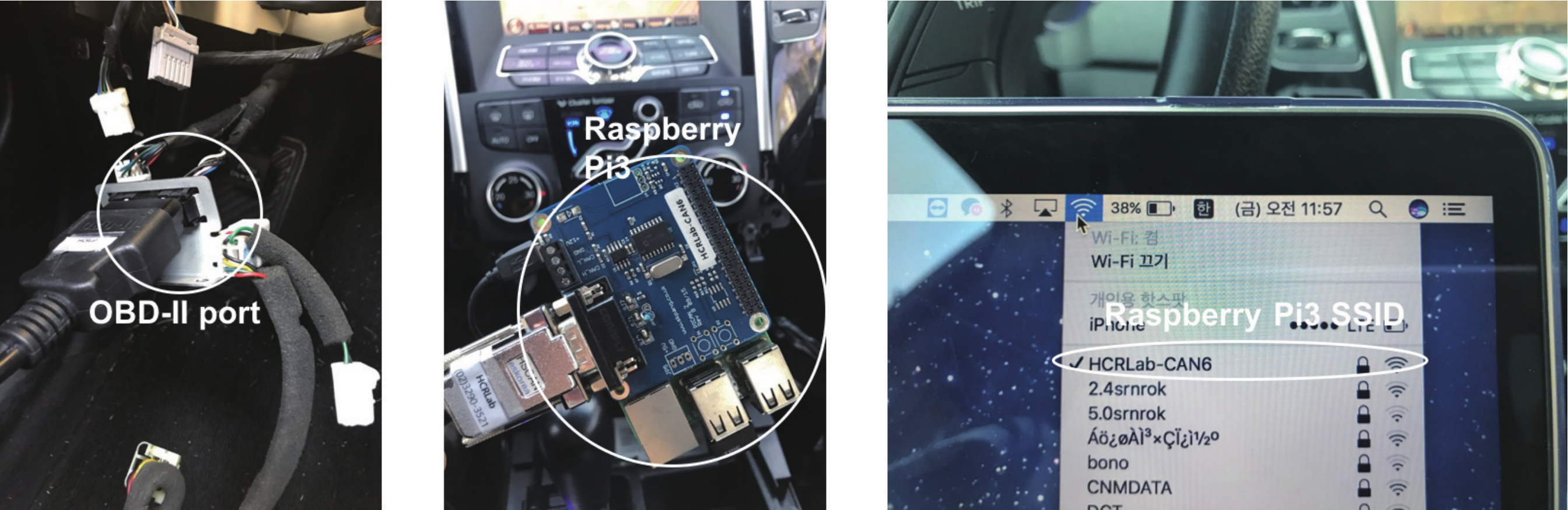}
	\caption{Data acquisition setup via OBD-II port of YF sonata with Raspberry Pi3}
	\label{fig:fig7}
\end{figure}

\subsection{Attack Design and Dataset}
Hyundai's YF Sonata is used as a testing vehicle. To capture CAN bus traffic, we plug Y-cable into OBD-II port; OBD-II port of YF Sonata is located under the steering wheel. Then, Raspberry Pi3 is used to connect to CAN bus. Also, a laptop computer is connected to Raspberry Pi3 through WiFi as shown in Fig. \ref{fig:fig7}. 

We launched four types of attacks on CAN bus as illustrated in Fig. \ref{fig:fig6}. Each attack is defined as follows. 

\begin{enumerate}
	\item DoS attack: Dos attack is to inject high priority of CAN messages (e.g. `0x000’ CAN ID packet) in a short cycle. We injected ‘messages of `0x000' CAN ID every 0.3 milliseconds. 
	\item FUZZY attack: Fuzzy attack is to inject messages of spoofed random CAN ID and DATA values. We injected messages of CAN ID and CAN data every 0.5 milliseconds. 
	\item RPM/GEAR attack: RPM/GEAR attack is to inject messages of certain CAN ID related to RPM/GEAR information. We injected messages related to RPM/GEAR every 1 millisecond.
\end{enumerate}

After data acquisition, we did labeling for the captured attack-free state traffic and attack traffic data. We released the dataset used in our experiments to foster further research. We make our dataset available at http://ocslab.hksecurity.net/Datasets/CAN-intrusion-dataset.  

Table \ref{table:table1} shows dataset which was used to test our model. The dataset consists of the training dataset and test dataset. The dataset was extracted from the running vehicle for about 10 minutes and contains both normal and abnormal packet with labeling. In Table \ref{table:table1}, `\#CAN message' means the total number of CAN packets including abnormal and normal ones during the attacks. `\#Attack image' means the only total number of the CAN images containing at least one abnormal CAN packet. Each dataset in Table \ref{table:table1} is independent of each other and not a multi-class.  

\begin{table}[!h]
	\caption{Data type and size}
	\label{table:table1}
	\centering
	\begin{tabular}{| c | c | c | c | }
		\hline
		\textbf{Data} & \textbf{Attack type} & \textbf{\#CAN message} & \textbf{\#Attack image}  \\ \hline
		Training set &   Normal data & 1,171,637 &  N/A  \\ \hline
		\multirow{4}{*}{Test set} 
		&    DoS attack data& 3,665,771 &  17,128  \\  \cline{2-4}
		&    FUZZY attack data& 3,838,860 &   20,317 \\   \cline{2-4}
		&    RPM attack data& 4,621,702 &  32,501 \\ \cline{2-4}
		&    GEAR attack data& 4,443,142&  29,751  \\ \hline
		
	\end{tabular}
\end{table}

\begin{figure*} [!h] 
	\centering
	\subfloat[binary image density]{
		\includegraphics[width=0.4\linewidth]{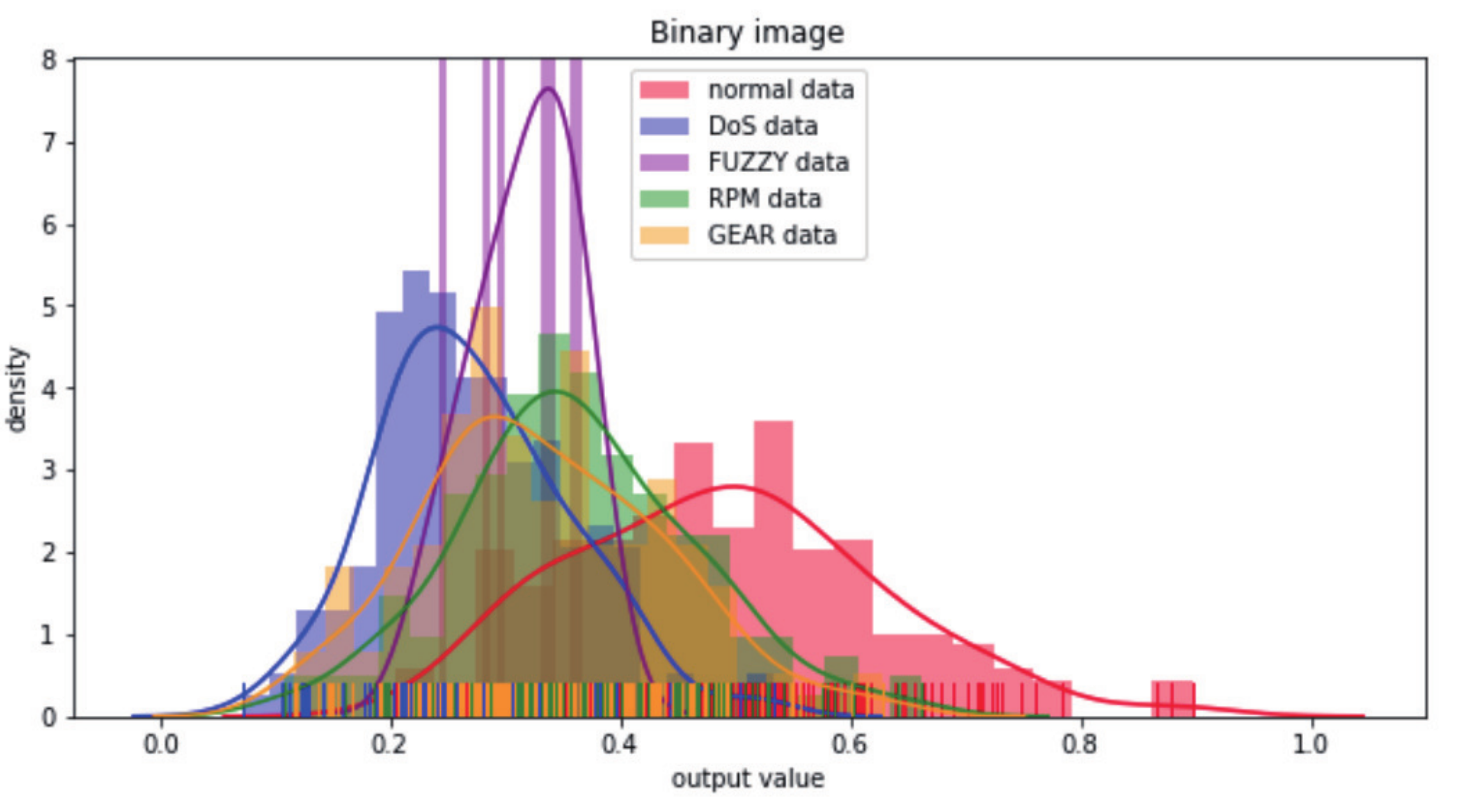}}
	\label{1a}\hfill
	\subfloat[\texttt{GIDS} image density]{
		\centering
		\includegraphics[width=0.41\linewidth]{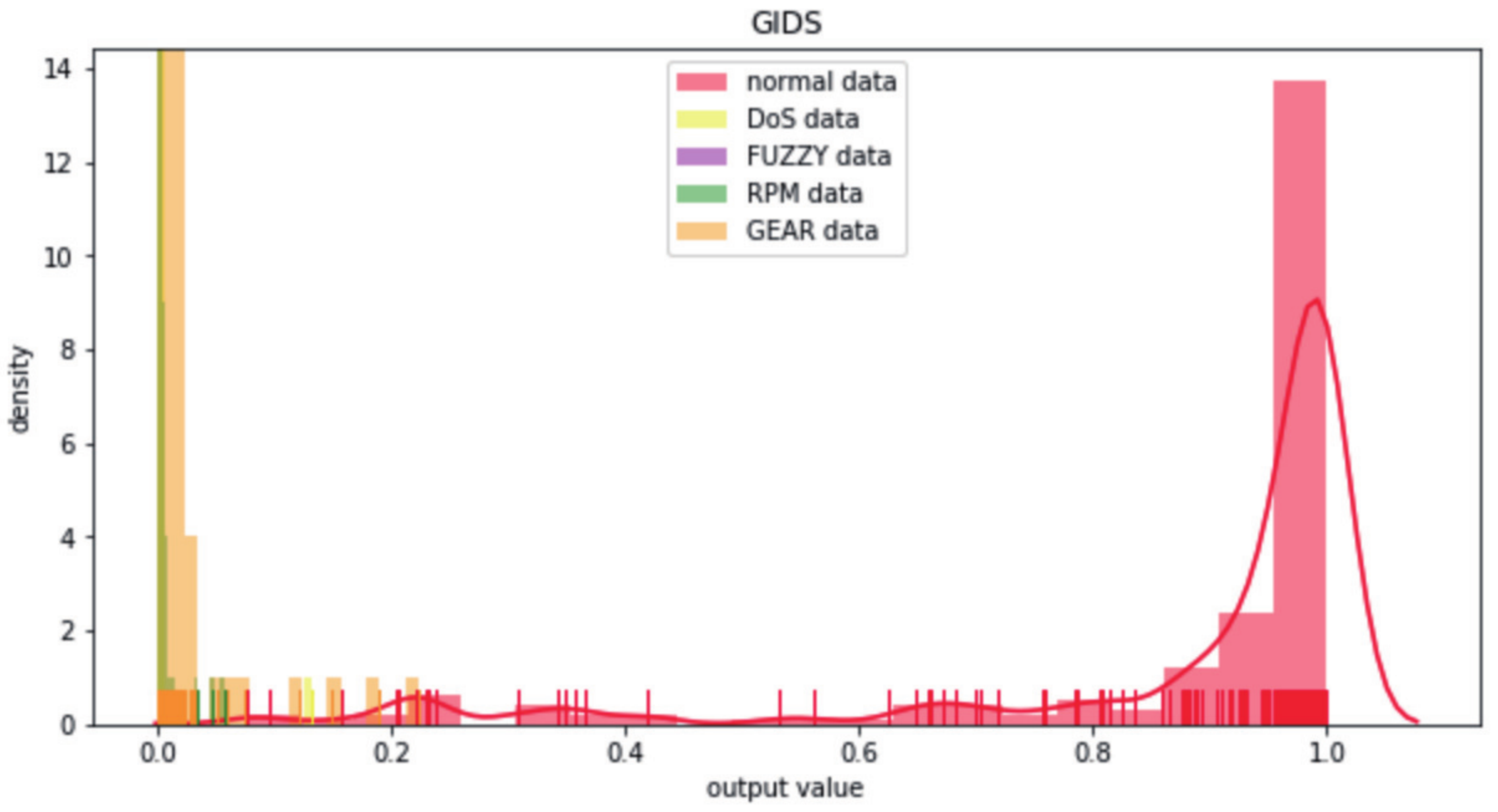}}
	\label{1b}\hfill
	\caption{Distribution of the output before and after one-hot-vector encoding}
	\label{fig:fig10} 
\end{figure*}

\subsection{Evaluating One-hot-vector Encoding}
\begin{figure}[!ht]
	\centering
	\parbox{1.2in}{
		\includegraphics[width=0.13\textwidth]{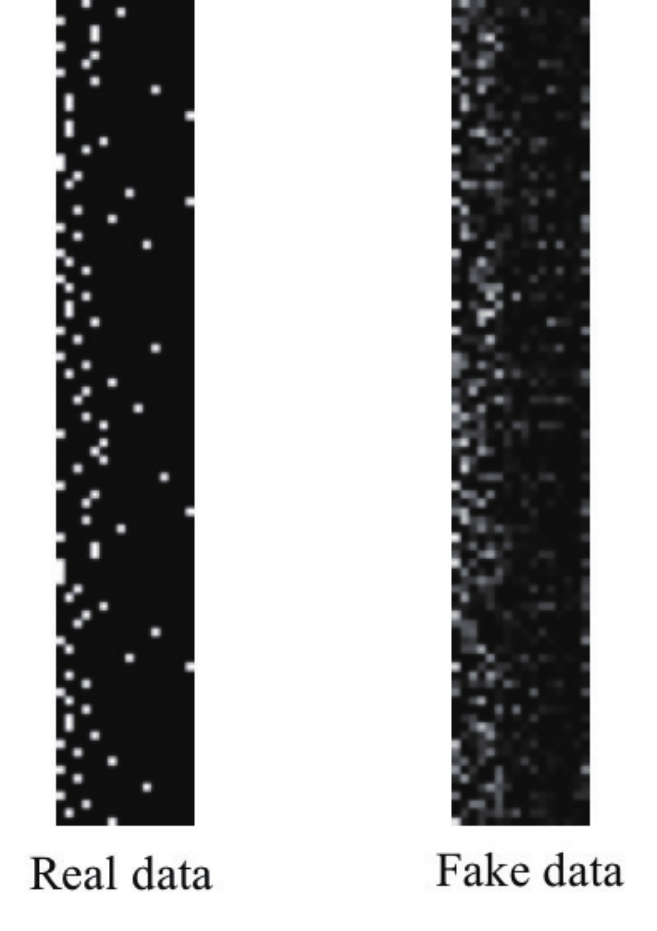}
		\caption{Image samples encoded with one-hot-vector}
		\label{fig:fig8}}
	\qquad
	\begin{minipage}{1.2in}
		\includegraphics[width=0.98\textwidth]{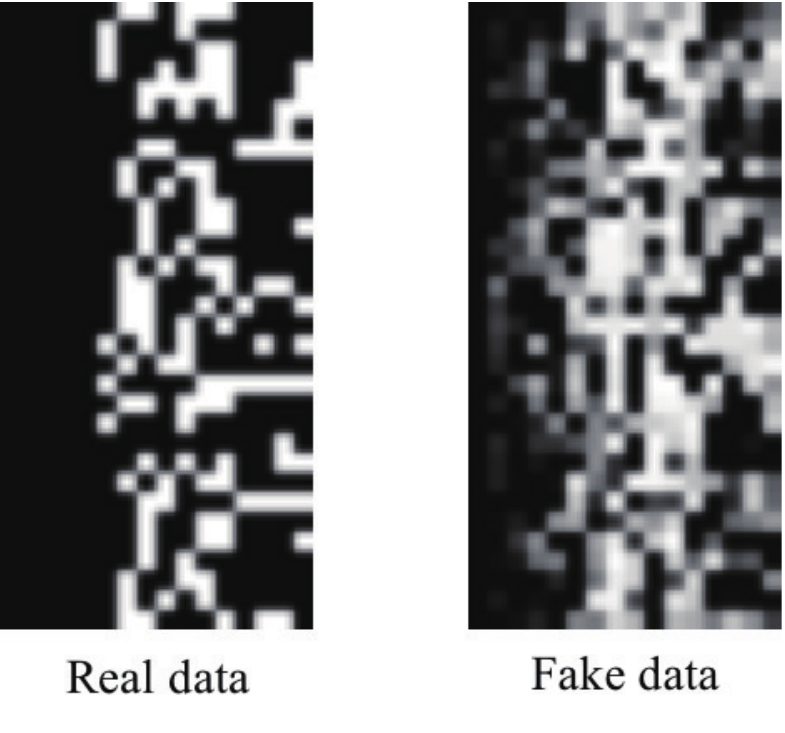}
		\caption{Image samples not encoded with one-hot-vector}
		\label{fig:fig9}
	\end{minipage}
\end{figure}

We converted the CAN data extracted from the vehicle into simple form images by encoding them with one-hot-vector. Fig. \ref{fig:fig8} shows the images generated by this way, and Fig. \ref{fig:fig9} shows the images not encoded with the one-hot-vector; All CAN IDs of 11 bits are converted into an image. In both Fig. \ref{fig:fig8} and Fig. \ref{fig:fig9}, the left image is real CAN image and the right image is fake CAN image generated by the generator. Fig. \ref{fig:fig8} shows more simple form whereas Fig. \ref{fig:fig9} shows complex form. 

Encoding with one-hot-vector can reduce the required time and show better performance than when the binary CAN data are converted as it is. Fig. \ref{fig:fig10} shows the distribution of the output of \texttt{GIDS} before and after one-hot-vector encoding. In the left model which uses binary images as it is, the output is widely distributed from 0 to 1. On the other hand, as shown in the right model, \texttt{GIDS} model using images converted with one-hot-vector has a certain threshold to classify normal data and abnormal data. That is, one-hot-vector encoding allows the intrusion detection model to separate normal data and attack data explicitly.

\subsection{Hyperparameters}
Based on the experiment, we set some hyperparameters of the \texttt{GIDS} model, which can improve the detection performance of the \texttt{GIDS} model. \texttt{GIDS} shows different performance according to values of these parameters. Parameters consist of detection threshold, attack threshold, and input size. We found the most suitable values of parameters through experiments and applied it to the final \texttt{GIDS} model. We present experimental results for each parameter as follows. 

\begin{enumerate}
	\item \textbf{Detection threshold}: The outputs of \texttt{GIDS} model are 0 to 1. Among these outputs, \texttt{GIDS} classify attack data and normal data by a specific detection threshold. We define detection threshold as 0.1. If the output of the \texttt{GIDS} model is less than 0.1, it is judged to an anomaly. Although some outputs in the normal data were distributed below 0.1 as in Fig. \ref{fig:fig10}(b), it may be regarded as an error that can appear in the sampling process.
	\item \textbf{Input size}: Input size means a unit to convert CAN IDs into images. CAN IDs extracted from the vehicle are grouped by input size and they are converted into images. We measured the accuracy of the \texttt{GIDS} model, increasing the input size from a minimum of 32 to a maximum of 128. Experimental results showed that the accuracy increased until 64 input size, but it tended to decrease after that as shown in Fig. \ref{fig:fig12}. Therefore, we define an input size as 64, and it can be changed flexibly depending on the vehicle environment.
	
	\begin{figure}[!ht] 
		\centering
		\includegraphics[width=0.6\textwidth]{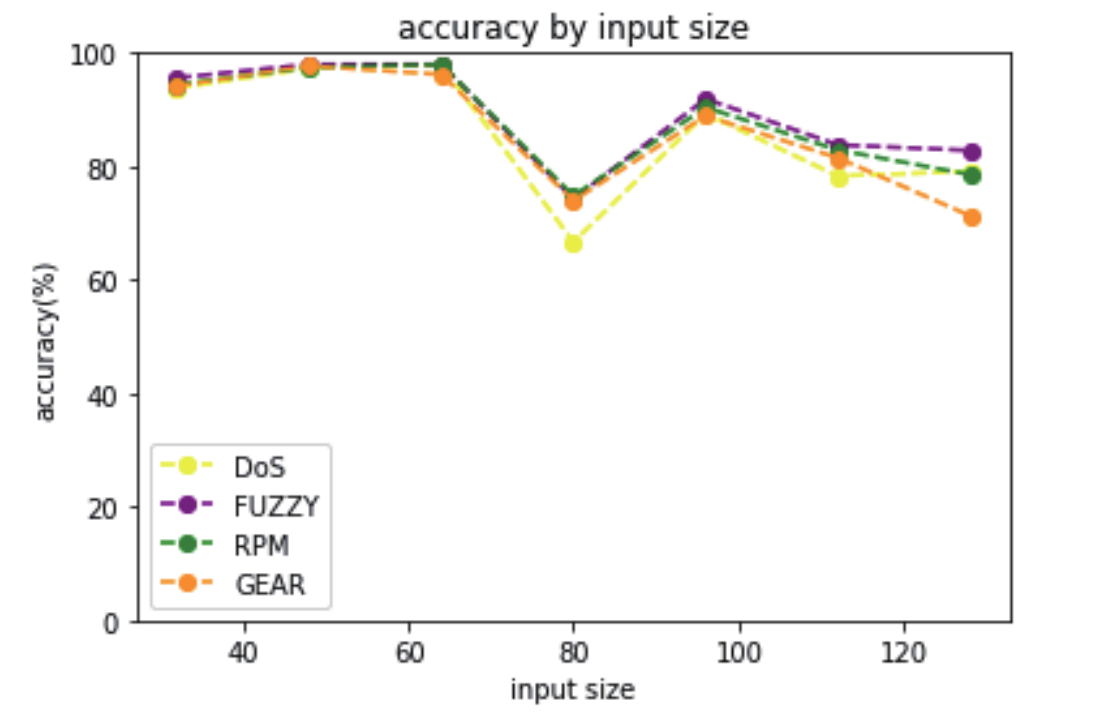}
		\caption{Accuracy of GIDS according to input size}
		\label{fig:fig12}
	\end{figure}

	\begin{table*}[!t]
		\caption{Detection rate of the first discriminator in GIDS according to attack data of training set}
		\centering
		\small\addtolength{\tabcolsep}{-3pt}
		\label{table:table3}
		\begin{tabular}{| c | c | c| c | c | c| c|}
			\hline \textbf{training set}  & \textbf{DoS detection rate} & \textbf{FUZZY detection rate} & \textbf{RPM detection rate} & \textbf{GEAR detection rate}  & \textbf{Normality detection rate}  \\ \hline
			DoS &  \textbf{99.9\%} & 0.0\% &0.0\% &0.0\% & 99.9\%  \\ \hline
			FUZZY   &2.0\% &  \textbf{98.7\%} &33.0\% &1.9\% &100.0\%  \\ \hline
			RPM  &0.0\%&0.0\% & \textbf{99.6}\% & 0.0\% &100.0\% \\ \hline
			GEAR  &0.0\%&0.9\%&0.0\% & \textbf{99.8\%} &99.6\% \\ \hline
		\end{tabular}
	\end{table*}
	
	\item \textbf{Attack threshold}: The attack threshold is a criterion for judging attack CAN images. We define attack threshold as 1. That is, if at least one attack packet is included in the CAN image, it is judged to be an abnormal image. We improve the security of the \texttt{GIDS} model by detecting even occasionally injected abnormal packets.
\end{enumerate}  

\subsection{Experiment Result}
Firstly, we tested the accuracy of the first discriminator which is trained using known attack data. Table \ref{table:table3} shows the detection rates of first discriminator for each attack data. As results of the experiment, attack data used in the training process were detected well but attack data not used for training were hardly detected. It requires a new detection model that can detect attacks even if only normal data are used in the training process. 

Secondly, we tested the detection accuracy of the second discriminator which uses random fake data in the training process instead of the real attack data. Table \ref{table:table2} shows the detection performance for each of the four attack data. Any attacks in the Table \ref{table:table2} were not used in the training process of the second discriminator. As results of the experiment, each of the four attacks was detected with an average of 98\% accuracy. Although the accuracy is less than 100\%, we can improve the accuracy of the \texttt{GIDS} model by combining it with first discriminator which uses attack data for the training process. 

\begin{table}[h]
	\caption{Performance of the second discriminator in \texttt{GIDS}}
	\centering
	\small\addtolength{\tabcolsep}{-3pt}
	\label{table:table2}
	\begin{tabular}{| c | c | c| c | c | c |}
		\hline
		\textbf{Data type} & \textbf{Detection rate} & \textbf{Precision} &\textbf{Accuracy} & \textbf{AUC}\\ \hline
		DoS attack&99.6\%&96.8\%&97.9\%&0.999\\ \hline
		FUZZY attack&99.5\%&97.3\%& 98.0\%&0.999\\ \hline
		RPM attack&99.0\%&98.3\%&98.0\%&0.999\\ \hline
		GEAR attack &96.5\%&98.1\%&96.2\%&0.996\\ \hline
	\end{tabular}
\end{table}

\section{Conclusion}
\label{conclusion}
In this study, we presented the \texttt{GIDS}, GAN based IDS for the in-vehicle network. Firstly, we proposed encoding a large number of CAN IDs with simple one-hot-vector, which can increase the performance and speed of the GIDS. Also, the proposed \texttt{GIDS} uses random fake data in the training process instead of the real attack data. It allows the \texttt{GIDS} model to detect unknown attacks with only normal data. Finally, we proposed a detection system that combines the first discriminator for detecting known attack data and the second discriminator for detecting unknown attack data. It can improve the detection accuracy of the proposed \texttt{GIDS} model. As a result of the experiment, The \texttt{GIDS} showed the average accuracy of 100\% for the first discriminator and the average accuracy of 98\% for the second discriminator. 

\texttt{GIDS} can be applied to the various types of the vehicle through the new training process and adjustment of the hyperparameters. Because the \texttt{GIDS} is pre-trained system and uses the deep-learning method, it is difficult to be manipulated by the attacker. Also, it can be real-time intrusion detection for the in-vehicle network. In practically, the number of messages that CAN bus system generates per second is about 1,954. \texttt{GIDS} takes only 0.18 seconds to detect about 1,954 CAN messages and it has a constant ratio of elapsed time for intrusion detection even if the amount of CAN data to be detected increases. 

The proposed \texttt{GIDS} model has the strengths of expandability, effectiveness, and security so it can be suitable IDS for the in-vehicle network.

\subsection{Discussion}
Although GAN based IDS describes the CAN network traffic well, it is still challenging point to distinguish anomalous traffic caused from `normal malfunctioning of electronic components' from anomalous traffic caused from `intentional attacks by hacker'. Nonetheless, GAN based IDS is still effective under the circumstance of lack of `known attack patterns for vehicles' such as nowadays. GAN based IDS and its evaluation becomes more precise as many attack patterns for vehicles become revealed.

\section*{Acknowledgment}
This work was supported by Institute for Information \& communications Technology Promotion(IITP) grant funded by the Korea government(MSIT) (No. R7117-16-0161, Anomaly Detection Framework for Autonomous Vehicles)

\bibliographystyle{unsrt}  
\bibliography{template}

\end{document}